\documentclass[a4paper,reqno]{amsart}
\usepackage[T1]{fontenc}
\usepackage[english]{babel}
\usepackage{amssymb,bbm,enumerate}
\usepackage{xcolor}
\usepackage[linktocpage=true,colorlinks=true, linkcolor=blue, citecolor=red, urlcolor=green]{hyperref}
\usepackage{multirow} 
\usepackage{graphicx}

\usepackage{float}
\restylefloat{table}
\usepackage{tabu}

\renewcommand{\phi}{\varphi}

\renewcommand{\Re}{\operatorname{Re}}

\newcommand{\bb}[1]{\mathbb{#1}}
\newcommand{\mc}[1]{\mathcal{#1}}
\newcommand{\mf}[1]{\mathfrak{#1}}

\newcommand{\beq}{\begin{equation}}
\newcommand{\eeq}{\end{equation}}
\newcommand{\e}{\varepsilon}

\newcommand{\de}{\partial}
\newcommand{\la}{\lambda}

\DeclareMathOperator{\ad}{ad}

\DeclareMathOperator{\op}{op}
\DeclareMathOperator{\Op}{Op}

\theoremstyle{plain}
\newtheorem{theorem}{Theorem}[section]

\newtheorem{proposition}[theorem]{Proposition}

\newtheorem{asu}{Assumption}

\theoremstyle{definition}

\theoremstyle{remark}
\newtheorem{remark}[theorem]{Remark}

\numberwithin{equation}{section}

\definecolor{light}{gray}{.9}

\usepackage{float}
\restylefloat{table}
\usepackage{mathtools}
\usepackage{tikz}
\usetikzlibrary{chains}

\tikzset{node distance=2em, ch/.style={circle,draw,on chain,inner sep=2pt},chj/.style={ch,join},every path/.style={shorten >=4pt,shorten <=4pt},line width=1pt,baseline=-1ex}

\usepackage{datenumber}

\newcounter{dateone}
\newcounter{datetwo}

\newcommand{\difftoday}[3]{%
\setmydatenumber{dateone}{\the\year}{\the\month}{\the\day}%
\setmydatenumber{datetwo}{#1}{#2}{#3}%
\addtocounter{datetwo}{-\thedateone}%
\textcolor{red}{\the\numexpr-\thedatetwo day(s) late}
}

\title{Opers for higher states of the quantum Boussinesq model}

\author{D.Masoero, A.Raimondo}

\address{Grupo de F\'isica Matem\'atica da Universidade de Lisboa,
Campo Grande, Lisboa, Portugal.}
\email{dmasoero@gmail.com}

\address{Dipartimento di Matematica e Applicazioni, Universit\'a di Milano-Bicocca,
Via Roberto Cozzi Marconi 55, I-20125 Milano, Italy}
\email{andrea.raimondo@unimib.it}
\begin{document}

\pagestyle{plain}

\begin{abstract} 
We study the ODE/IM correspondence for all the states of the quantum Boussinesq model.
We consider a particular class of third order linear ordinary differential operators and show
that the generalised monodromy data of such operators provide solutions to the Bethe Ansatz equations
of the Quantum Boussinesq model. 
\end{abstract}

\maketitle

%\tableofcontents

%%%
\section{Introduction}
The quantum Boussinesq model \cite{bazhanov02integrable} is a 2 dimensional
conformal field theory with a $\mathcal{W}_3$ symmetry, and it can be exactly solved via the Bethe Ansatz equations.
This model can be realised as the quantisation of a $\mf{sl}_3$ Drinfeld-Sokolov hierarchy, or as the continuum limit
of a $\mf{sl}_3$ XXZ chain.
It belongs to a large family of theories which are known as $\mf{g}-$quantum KdV models; they exist for any Kac Moody algebra
$\mf{g}$ \cite{feigin96} (in
the present case $\mf{g}=\widehat{\mf{sl}}_3$), and in the simplest case, namely $\mf{g}=\widehat{\mf{sl}}_2$, the Hamiltonian
structure of such a theory is the quantisation of the second Poisson structure of the classical KdV equation \cite{baluzaI}.

According to the celebrated ODE/IM correspondence \cite{doreytateo98,dota00,bazhanov01,BLZ04,dorey07,doreyreview,FF11,marava15,marava17,mara18,kotousov2019bethe}
to every state of the $\mf{g}$ quantum KdV model there corresponds a unique $\mf{g}^L$ oper (here $\mf{g}^L$ is the Langlands dual
of $\mf{g}$) whose generalised monodromy data provide the solution of the Bethe Ansatz equations of that state.

In our previous paper \cite{mara18} we constructed the opers  corresponding to higher states of the
$\mf{g}$ quantum KdV model, for any $\mf{g}$ untwisted affinization of a simply laced Lie algebra. This was done by following the definition given in \cite{FF11};  solutions to the Bethe Ansatz were obtained based on our previous works \cite{marava15,marava17}.

In this note we provide explicit and simpler formulas for opers corresponding to higher states of the quantum Boussinesq model,
by specialising the results of \cite{mara18} to the case $\mf{g}=\widehat{\mf{sl}}_3$. This serves two purposes:
we illustrate the general theory and its somehow heavy machinery
in terms of familiar and simple objects, and we find formulas which are much closer to the original work on higher states opers of the $\widehat{\mf{sl}_2}-$quantum KdV model \cite{BLZ04}, where higher states are conjectured to correspond to Schr\"odinger operators with a \textit{monster potential}.

%:
As the result of the present paper, we conjecture that the level $N$ states of the quantum Boussinesq model correspond
to the following third order differential operators:
\begin{align}\nonumber
L&=\de_z^3-\left(\sum_{j=1}^N\left(\frac{3}{(z-w_j)^{2}}+\frac{k}{z(z-w_j)}\right)+\frac{\bar{r}^1}{z^{2}}\right)\de_z\\ \label{eq:a2scalaroper}
&+\sum_{j=1}^N\left(\frac{3}{(z-w_j)^{3}}+\frac{a_j}{z(z-w_j)^{2}}+\frac{2(k+3)a_j-k^2}{3z^2(z-w_j)}\right)
+\frac{\bar{r}^2}{z^{3}}+\frac{1}{z^{2}}+\lambda z^k,
\end{align}
where $-3<k<-2$, and $\bar{r}^1,\, \bar{r}^2\in\bb{C}$, and
where the $2N$ complex variables $\{a_\ell,w_\ell\}_{\ell=1,\dots,N}$, satisfy the following system of $2N$ algebraic equations
\begin{subequations}\label{monodromy}
\beq\label{monodromy1}
a_\ell^2-ka_\ell+k^2+3k-3\bar{r}^1=\sum_{\substack{j=1,\dots,N\\ j\neq \ell}}\left(\frac{9w_\ell^2}{(w_\ell-w_j)^2}+\frac{3kw_\ell}{w_\ell-w_j}\right),
\eeq
\begin{align}
Aa_\ell+B-9(k+2)w_\ell=\sum_{\substack{j=1\\ j\neq \ell}}^N&\left(\frac{18(k-a_\ell-a_j)w_\ell^3}
{(w_\ell-w_j)^3}+\frac{(12k+9k^2-(63+6k)a_j-9ka_\ell)w_\ell^2}{(w_\ell-w_j)^2}\right.\notag\\
&+\left.\frac{(9k+16k^2+6(k^2+10k+6)a_j-5ka_\ell)w_\ell}{w_\ell-w_j}\right).\label{monodromy2}
\end{align}
\end{subequations}
The parameters $A,B$ are given by
\begin{align*}
A&=14k^2+50k-8\bar{r}^1+45,\\
B&=27(\bar{r}^1-\bar{r}^2)-k(7k^2+7k+9\bar{r}^2-13\bar{r}^1+9),
\end{align*}
and the additional singularities $w_j$, $j=1\dots,N$ are assumed to be pairwise distinct and nonzero.
The system of algebraic equation \eqref{monodromy}
is equivalent to the requirement that the monodromy around the singular point $z=w_j$ is trivial for all $j=1\dots N$,
independently on the parameter $\lambda$.

The correspondence among the free parameters $\la,\bar{r}^1,\bar{r}^2,k$ of the above equations and the 
the free parameters $c,(\Delta_2,\Delta_3), \mu$ (respectively the central charge, the highest weight, the spectral parameter)
of the Quantum Boussinesq model, as constructed  in \cite{bazhanov02integrable} (more about this below), goes as follows:
\begin{subequations}\label{parametersintro}
 \begin{align}\label{eq:ua}
c&=-\frac{3(4k+9)(3k+5)}{k+3},\\ \label{eq:ub} 
\Delta_2&=\frac{(\bar{r}^1-8)k^2+6(\bar{r}^1-5)k+9\bar{r}^1-27}{9(k+3)}\\ \label{eq:uc}
\Delta_3&=\frac{(k+3)^{3/2}}{27}(\bar{r}^1-\bar{r}^2) \\ \label{eq:corrparameters}
\la&=-i\,\Gamma(-k-2)^3\mu^3,
\end{align}
\end{subequations}
where $\Gamma(s)$ denotes the $\Gamma$ function with argument $s$. Moreover
the integer $N$, which is the number of additional regular singularities in \eqref{eq:a2scalaroper}, coincides with the level of the state.
Hence, system \eqref{monodromy} is expected to possess
$p_2(N)$ solutions, where $p_2(N)$ is the number of bi-coloured partitions of $N$.

The paper is organised as follows. In Section \ref{sec:Qopers}
we introduce the quantum KdV opers, following \cite{mara18} (which in turns builds on \cite{FF11}), and derive from the general theory of the formulas
\eqref{eq:a2scalaroper}, \eqref{monodromy}. In Section \ref{sec:BA} we review
the construction of solutions of the Bethe Ansatz equations as generalised monodromy data,
following \cite{marava15,mara18}. Finally, in Section \ref{sec:QBM} we briefly summarise 
the construction of the quantum Boussinesq model provided in \cite{bazhanov02integrable}. 

This work deals with differential equations and representation theory.
We omit many proofs of the analytic results, which can be found in greater generality in \cite{mara18}. However, we do provide all details
of the algebraic calculations.

\subsection*{Acknowledgements}
The authors are partially supported by the FCT Project PTDC/MAT-PUR/ 30234/2017 `Irregular
connections on algebraic curves and Quantum Field Theory'. D. M. is supported by the
FCT Investigator grant IF/00069/2015 `A mathematical framework for the ODE/IM correspondence'.
The authors thank Dipartimento di Matematica dell'Universit\`a degli studi di Genova for the kind hospitality.

%%%
\section{Quantum KdV Opers}\label{sec:Qopers}
In this Section we introduce the Quantum KdV opers, as defined in \cite{FF11}, in the special case
$\mf{g}=\widehat{\mf{sl}}_3$, and derive the third order scalar differential operator \eqref{eq:a2scalaroper}. The reader should refer to \cite{mara18}, and references therein for more details.

We begin by introducing some theory of the
algebra $\mf{sl}_3(\bb{C})$\footnote{For sake of simplicity we prefer to work with $\mf{sl}_3$-opers, instead of $\widehat{\mf{sl}}_3$-opers.
We do that by considering the loop algebra variable $\la$ as a free complex parameter. More about this in \cite[Section 4]{mara18}.},
which we realise as the Lie algebra of traceless 3 by 3 matrices
(in such a way that it coincides with its first fundamental representation, also known as standard representation).
The algebra has the decomposition $\mf{n}_-\oplus \mf{h} \oplus \mf{n}_+$,
where $\mf{n}_-$ are lower diagonal matrices, $\mf{h}$ is the Cartan subalgebra  of traceless diagonal matrices, and
$\mf{n}_+$ are upper diagonal matrices. The subalgebra $\mf{b}_+:=\mf{h}\oplus \mf{n}_+$ is called the Borel subalgebra. We provide an explicit basis of $\mf{b}_+$ as follows

\begin{equation}h_1=
\begin{pmatrix}
1 & 0 & 0\\
0 & -1 & 0\\
0 & 0 & 0
\end{pmatrix} \quad h_2=\begin{pmatrix}
0 & 0 & 0\\
0 & 1 & 0\\
0 & 0 & -1
\end{pmatrix}
\end{equation}

\begin{equation}
e_1=
\begin{pmatrix}
0 & 1 & 0\\
0 & 0 & 0\\
0 & 0 & 0
\end{pmatrix} \quad 
e_2=
\begin{pmatrix}
0 & 0 & 0\\
0 & 0 & 1\\
0 & 0 & 0
\end{pmatrix} \quad
e_{\theta}=
\begin{pmatrix}
0 & 0 & 1\\
0 & 0 & 0\\
0 & 0 & 0
\end{pmatrix} 
\end{equation}
We introduce three further elements, the sum of the negative Chevalley generators of the Lie algebra $f \in \mf{n}_-$ (principal nilpotent element), the dual of the Weyl vector $\rho^\vee \in \mf{h}$, and the dual of the highest root $\theta^\vee \in \mf{h}$. We have:
\begin{equation}
f=
\begin{pmatrix}
0 & 0 & 0\\
1 & 0 & 0\\
0 & 1 & 0
\end{pmatrix} \quad 
\rho^\vee=\theta^\vee=
\begin{pmatrix}
1 & 0 & 0\\
0 & 0 & 0\\
0 & 0 & -1
\end{pmatrix} 
\end{equation}
The unipotent group $\mc{N}= \lbrace \exp{y}, y \in \mf{n}_+ \rbrace$ acts on $\mf{sl}_3$ via the formula
\begin{equation*}
\exp{y}.g=g+\sum_{k\geq1}\frac{(\ad_y)^k.g}{k!}, \qquad \ad_y.g:=[y,g] , 
\end{equation*}
and the affine subspace $f+\mf{b}_+$ is preserved by the action.
Following Kostant \cite{Ko78}, and given a vector subspace $\mf{s}\subset\mf{n}_+$, we say that the affine subspace $f+\mf{s}$ is a \emph{transversal space} if
\begin{enumerate}
 \item The orbit of $f+\mf{s}$ under the action of $\mc{N}$ coincides with $f+\mf{b}_+$
 \item For each $s \in \mf{s}$, then $\exp{y}. (f+s) \notin f+\mf{s}$ unless $y=0$
\end{enumerate}
The subspace $\mf{s}=\bb{C}e_1\oplus \bb{C}e_{\theta}$
satisfies the above hypotheses\footnote{As an example, the Cartan subalgebra $\mf{h}$ satisfies the first but not the second hypothesis above} and
the transversal space $f+\mf{s}$ is the space of companion matrices:
$$f+\mf{s}=\left\{\begin{pmatrix} 0 & a & b \\ 1 &0 & 0\\ 0& 1& 0\end{pmatrix}\,|\,a,b\in \bb{C}\right\}.$$
We fix this choice for the rest of the paper.

\subsection{Opers}
We denote by $K$ the field of rational functions in the variable $z$, and we define
\begin{enumerate}
 \item $\mf{g}(K),\mf{b}_+(K),\mf{n}_+(K)$ the Lie algebras of rational functions with values in
$\mf{g},\mf{b}_+,\mf{n_+}$ respectively.
\item The space of (global meromorphic) $\mf{g}-$valued connections
$\mbox{conn}(K)=\lbrace \partial_z+ g, g \in \mf{g}(K) \rbrace$.
\item The subset  $\op(K)=\lbrace \mc{L}= \partial_z + f+ b, b\in \mf{b}_+(K)\rbrace\subset\mbox{conn}(K) $.
\item The group of unipotent Gauge transformations $\mc{N}(K)=\lbrace \exp y, y \in \mf{n}_+(K) \rbrace$,
acting on $\mbox{conn}(K)$ via the  formula
\begin{equation}
 \exp{y}.(\partial_z+g)=\partial_z-\sum_{k\geq0}\frac{1}{(k+1)!}(\ad_{y})^k\frac{d y}{dz}+ \exp{y}.g.
\end{equation}
Note that the above action preserves the subset $\op(K)$.
\item The space of $\mf{sl}_3$ opers as $\Op(K)=\op(K)/\mc{N}(K)$.
\end{enumerate}
The space of opers $\Op(K)$ admits a very explicit description once a transversal space $f+\mf{s}$ is fixed: any element in
$\op(K)$ is Gauge equivalent to a unique connection of the form $\partial_z+f+s, s \in \mf{s}(K)$. Hence we have a bijection
$$\Op(K)\cong \lbrace \partial_z+f+s,  s \in \mf{s}(K) \rbrace.$$ 
We call $\partial_z+f+s$ the \emph{canonical form} of any oper Gauge equivalent to it.

\subsection{Opers and scalar ODEs}
It is a standard and elementary result that the space of $\mf{sl}_3$ opers coincides with the space
of third order linear scalar differential operators (with principal symbol equal to $1$ and vanishing sub-principal symbol). Indeed, for what we have said so far,
any oper has a unique representative of the form 
$$\mc{L}=\partial_z+f+v_1(z)e_1+v_2(z)e_\theta,$$
where $v_1,v_2$ are a pair of (arbitrary) rational functions. In the fisrt fundamental representation, this oper takes the form
\beq\label{opercanmatrix}
\mc{L}=\partial_z+\begin{pmatrix} 0 & v_1(z) & v_2(z) \\ 1 &0 & 0\\ 0& 1& 0\end{pmatrix}.
\eeq
If $\{\epsilon_1,\epsilon_2,\epsilon_3\}$ is the standard basis of $\bb{C}^3$, and given $\psi= \bb{C} \to \bb{C}^3$, with $\psi(z)=\psi_1(z)\epsilon_1+\psi_2(z)\epsilon_2+\psi_3(z)\epsilon_3$, then  the matrix first order equation
$$ \mc{L} \psi(z)=0,$$ 
is easily seen to be equivalent to the following scalar ODE for the third coefficient $\Psi:=\psi_3$
\begin{equation}\label{eq:scalarform}
(\partial_z^3- v_1 \partial_z + v_2 ) \Psi(z)=0.
\end{equation}
We will use this scalar representation in the rest of the paper.

\subsection{(Ir)Regular Singularities}
Let $\mc{L}$ be an oper in the canonical form \eqref{opercanmatrix}, and $w\in \bb{C}$ a pole of $v_1$ or $v_2$, so that
\begin{align*}
v_1&=\bar{s}_1 (z-w)^{-\delta_1}+o((z-w)^{-\delta_1}),\\
v_2&=\bar{s}_2 (z-w)^{-\delta_2} +o((z-w)^{-\delta_2})
\end{align*}
for some $\bar{s}^1,\bar{s}^2 \neq 0$ and some $\delta_1,\delta_2\in\bb{Z}$. We define \cite{mara18}
\begin{itemize}
 \item  The \emph{slope} of the singular point $w\in\bb{C}$ as 
 $$\mu=\max\left\{1,\max\left\{\frac{\delta_1}{2},\frac{\delta_2}{3}\right\}\right\} \in \bb{Q}$$
 \item  The \emph{principal coefficient} of the singular point $w$ as 
 $$f-\rho^\vee+\bar{s}^1e_1+\bar{s}^2e_2=\begin{pmatrix} -1 & \bar{s}^1 & \bar{s}^2 \\ 1 &0 & 0\\ 0& 1& 1\end{pmatrix}\qquad \text{if}\qquad \mu=1,$$
and
 $$\quad f+\bar{s}^1e_1+\bar{s}^2e_2= \begin{pmatrix} 0 & \bar{s}^1 & \bar{s}^2 \\ 1 &0 & 0\\ 0& 1& 0\end{pmatrix}\qquad \text{if}\qquad \mu>1.$$
 \end{itemize}
As proved in \cite{mara18}, the singularity is \emph{regular} (in the sense of linear connections) if $\mu=1$ and \emph{irregular} if $\mu>1$.

\begin{remark}
In the case when  $w=\infty$, we write $v_1=z^{\delta_1}+o(z^{\delta_1})$, and $v_2=\bar{s}_2 z^{\delta_2} +o(z^{\delta_2})$
for some $\bar{s}^1,\bar{s}^2 \neq 0$, and $\delta_1,\delta_2\in\bb{Z}$, and define the slope of $w=\infty$ as  $\mu=\max\lbrace1,\max\lbrace\frac{\delta_1}{2},\frac{\delta_2}{3}\rbrace+2\rbrace$. The principal coefficient is defined as above.
\end{remark}

\subsection{$\mf{sl}_3-$quantum KdV Opers}
We define $\mf{sl}_3-$quantum KdV opers following \cite{FF11}. To this aim we fix
$-3<k<-2$ and $\bar{r}^1,\bar{r}^2 \in \bb{C}$ and write
\begin{equation}\label{eq:quantumopers}
 \mc{L}(z,\lambda)=\mc{L}_{G,\mf{s}}(z,\lambda)+ s(z), \quad s \in K(\mf{s}) \; .
\end{equation}
Here $\mc{L}_{G,\mf{s}}$ is the ground state oper
 \begin{equation}\label{eq:LGs}
\mc{L}_{G,\mf{s}}(z,\la)=\partial_z+
\begin{pmatrix}
0 &  \bar{r}^{1}/z^{2} & \bar{r}^{2}/z^{3}+z^{-2}+\lambda z^{k}\\
1 & 0 & 0\\
0 &1 &0
\end{pmatrix}
\end{equation}
We notice that $\mc{L}_{G,\mf{s}}(z,\la)$ has two singular points: $z=0$ is a regular singularity with principal coefficient
$$\begin{pmatrix} -1 & \bar{r}^1 & \bar{r}^2 \\ 1 &0 & 0\\ 0& 1& 1\end{pmatrix},$$
while $z=\infty$ is an irregular singularity, with slope $\mu=\frac43$ and principal coefficient 
$$\begin{pmatrix} 0 & 0 & 1 \\ 1 &0 & 0\\ 0& 1& 0\end{pmatrix}.$$
As it will be reviewed in the next section, one can obtain solutions of the Bethe Ansatz equations
by considering the differential equation $\mc{L}_{G,\mf{s}}\psi=0$: more precisely these are obtained as coefficients of 
the expansion of the subdominant solution at $+\infty$ in terms of a distinguished basis of solutions defined at $z=0$.

In \cite{BLZ04}, Bazhanov, Lukyanov and Zamolodchikov proved that in the case $\mf{g}=\widehat{\mf{sl}}_2$, the ground state oper could be modified without altering the above global structure, so that the modified equations yield (different) solutions of the same Bethe Ansatz equations (as coefficients of the same expansion). Feigin and Frenkel \cite{FF11}  extended these idea to the case of a general Kac-Moody algebra, and conjectured that the higher level opers could be uniquely specified by imposing on the $\mf{s}-$valued function $s$ the $4$ conditions below. These condition were shown to sufficient \cite{mara18}, and are expected to be necessary for generic values of the parameters $\hat{k},\bar{r}^1,\bar{r}^2$ \cite{FF11}.
We say that the oper $\mc{L}(z,\la)$ of the form \eqref{eq:quantumopers} is a $\mf{sl}_3$-quantum KdV oper if it satisfies the following
4 assumptions:
\begin{asu}\label{asu1}
The slope and principal coefficient at $0$ do not depend on $s$.
\end{asu}
\begin{asu}\label{asu2}
The slope and principal coefficient at $\infty$ do not depend on $s$. 
\end{asu}
\begin{asu}\label{asu3}
All additional singular points are regular and the corresponding principal coefficients are conjugated
to the element $f-\rho^\vee-\theta^\vee \in f+\mf{h}$.
\end{asu}
\begin{asu}\label{asu4}
All additional singular points have trivial monodromy for every $\la \in \bb{C}$.
\end{asu}
The following proposition, which is Proposition 4.7 in \cite{mara18} specialised to the case
of $\mf{g}=\mf{sl}_3$, is a first characterisation of the Quantum KdV opers; it shows that they have the form
(\ref{eq:a2scalaroper}).
\begin{proposition}\label{pro:quasinormal}
An operator $\mc{L}(z,\la)$ of the form \eqref{eq:quantumopers} satisfies the first three Assumptions if and only if
there exists a (possibly empty) arbitrary finite
collection of non-zero mutually distinct complex numbers $\lbrace w_j \rbrace_{j \in J}\subset \bb{C}^\times$ and a
collection  of numbers $ \{a_{11}^{(j)},a_{21}^{(j)}, a_{22}^{(j)}\}_{j\in J}\subset \bb{C}$, such that $\mc{L}(z,\la)$ has the form
\beq \label{eq:3ass}
 \mc{L}(z,\la)=  \partial_z+
 \begin{pmatrix}
 0 & W_1 & W_2\\
 1 & 0 & 0\\
 0 & 1 & 0
 \end{pmatrix},
\eeq
where
\begin{subequations}\label{w1w2}
\begin{align}
W_1(z)&=\frac{\bar{r}^{1}}{z^{2}} +\sum_{j \in J}\left(\frac{3}{(z-w_j)^{2}}+\frac{a_{11}^{(j)}}{z(z-w_j)}\right) ,\\
W_2(z,\la)&=\frac{\bar{r}^{2}}{z^{3}} +\frac{1}{z^2}+\lambda z^k +\sum_{j \in J}\left(\frac{3}{(z-w_j)^{3}}+
\frac{a_{21}^{(j)}}{z(z-w_j)^{2}}+\frac{a_{22}^{(j)}}{z^2(z-w_j)}\right).
\end{align}
\end{subequations}
\end{proposition}
Note that when $J$ is empty than \eqref{eq:3ass} reduces to the ground state oper \eqref{eq:LGs}. If $J$ is not empty, than we set $J=\{1,\dots,N\}$, for some $N\in\bb{Z}_+$. In order to fully characterise the $\mf{sl}_3-$quantum KdV opers, we must impose the fourth and last Assumption on the opers of the form \eqref{eq:3ass}, namely the triviality of  the monodromy about all the additional singularities $w_j,j=1\dots N$. We notice that the opers of the form \eqref{eq:3ass} depend on the $4N$ complex parameters $\{a_{11}^{(j)},a_{21}^{(j)}, a_{22}^{(j)},w_j\}_{j \in1 \dots N}$.
We will show in the following subsection that the trivial monodromy conditions are equivalent
to a complete system of $4N$ algebraic equations, which in turn are equivalent to  \eqref{eq:a2scalaroper},\eqref{monodromy}.

\subsection{Trivial monodromy conditions}
We fix $\ell \in 1\dots N$ and study under which conditions the oper $\mc{L}(z,\la)$ of the form \eqref{eq:3ass} has trivial monodromy about
$w_{\ell}$.
As we showed in \cite{mara18}, Assumption \ref{asu3} (more precisely, the fact that $\theta^\vee$ is  a co-root) implies that the monodromy about $w_{\ell}$ is trivial if and only if it is trivial in at least one irreducible (nontrivial) representation.
In other words, it is necessary and sufficient that the monodromy at $z=w_\ell$ is trivial for the solutions of the equation
$\mc{L}(z,\la)\Psi=0$ in the standard representation.

To this aim we write the above equation in the scalar form
\begin{equation}\label{eq:scalarfrob}
 (\partial_z^3-W_1 \partial_z + W_2) \Phi(z)=0 \, ,
\end{equation}
and use the method of the Frobenius expansion, that is we look for solutions of the form
\begin{equation}\label{eq:frob}
\Phi^{(\beta)}(z)=\sum_{m\geq 0}\Phi_m(z-w_\ell)^{\beta+m}
\end{equation}
Writing the Laurent expansion of \eqref{eq:scalarfrob} at $w_{\ell}$ as
\begin{subequations}\label{eq:laurent}
\begin{align}
W_1(z)&=\sum_{m=0}^{+\infty}q_{1m}^{(\ell)}(z-w_\ell)^{m-2} ,\qquad  q_{10}^{(\ell)}=3, \\ 
W_2(z)&=\sum_{m=0}^{+\infty}q_{2m}^{(\ell)}(z-w_\ell)^{m-3} ,\qquad  q_{20}^{(\ell)}=3,
\end{align}
\end{subequations}
expanding the equation \eqref{eq:frob} in powers of $z-w_\ell$, and equating to zero term-by-term we obtain
\begin{align}\nonumber
& \Phi^{(\beta)}_0 P(\beta)=0 \\ \label{expansion2}
& P(\beta+r)\Phi^{(\beta)}_r=-\sum_{i=1}^n(-1)^i
\sum_{m=1}^r q_{im}^{(\ell)}\Phi^{(\beta)}_{r-m}\prod_{s=0}^{n-i-1}(\beta+r-m-s),
\end{align}
where the indicial polynomial $P(\beta)=(\beta-3)(\beta-1)(\beta+1)$. The roots of the indicial polynomial, $\beta=-1,1,3$, are known as \emph{indices}. Since the indices are integers, the monodromy matrix has a unique eigenvalue, $1$, with algebraic multiplicity $3$,
and the monodromy is trivial if and only if the recursion \eqref{expansion2} has a solution for all the indices. Indeed, in such a case,
$\Phi^{\beta}(e^{2\pi i}z)=\Phi^{\beta}(z)$ for $\beta=-1,1,3$; otherwise logarithmic terms must be added to the series \eqref{eq:frob}
and the monodromy is not diagonalizable \cite{wasowAs}.

We analyse the recursion \eqref{expansion2} separately for the three indices.

The recursion \eqref{expansion2} for the index $\beta=3$ admits always a unique solution, since $\Phi(3+r) \neq 0, \forall r\geq 1$.

In the case $\beta=1$, we have that $P(\beta+r)=0, r\geq 1$ if and only if $r=2$. Hence 
the recursion is over-determined.
Computing the first two terms we obtain
\begin{align*}
-3\Phi^{(1)}_1&= \left(q_{11}^{(\ell)}- q_{21}^{(\ell)}\right)\Phi^{(1)}_{0},\\
0 \times \Phi^{(1)}_2&= \left(2q_{11}^{(\ell)}- q_{21}^{(\ell)}\right)\Phi^{(1)}_{1}+
\left(q_{12}^{(\ell)}- q_{22}^{(\ell)}\right)\Phi^{(1)}_{0}.
\end{align*}It follows that the recursion for the index $\beta=1$ has at least one solution if and only if
\begin{equation}\label{eq:180117}
q_{12}^{(\ell)}- q_{22}^{(\ell)}=\frac{2}{3}\left(q_{11}^{(\ell)}\right)^2-
q_{11}^{(\ell)}q_{21}^{(\ell)}+\frac{1}{3}\left(q_{21}^{(\ell)}\right)^2
\end{equation}
Finally, the Frobenius method for the index $\beta_2=-1$ gives
\begin{align*}
3\Phi^{(-1)}_1&= -\left(q_{11}^{(\ell)}+q_{21}^{(\ell)}\right)\Phi^{(-1)}_{0},\\
0 \times \Phi^{(-1)}_2&= - q_{21}^{(\ell)}\Phi^{(-1)}_{1}-\left( q_{12}^{(\ell)}+ q_{22}^{(\ell)}\right)\Phi^{(-1)}_{0},\\
-3\Phi_3^{(-1)}&= \left(q_{11}^{(\ell)}- q_{21}^{(\ell)}\right)\Phi^{(-1)}_{2}- q_{22}^{(\ell)}\Phi^{(-1)}_{1}-
\left(q_{13}^{(\ell)}+q_{23}^{(\ell)}\right)\Phi^{(-1)}_{0},\\
0 \times \Phi^{(-1)}_4&= \left( 2q_{11}^{(\ell)}-q_{21}^{(\ell)}\right)\Phi^{(-1)}_{3}+
\left( q_{12}^{(\ell)}-q_{22}^{(\ell)}\right)\Phi^{(-1)}_{2}
-q_{23}^{(\ell)}\Phi^{(-1)}_{1}-
\left( q_{14}^{(\ell)}+q_{24}^{(\ell)}\right)\Phi^{(-1)}_{0},
\end{align*}
and we obtain the following constraints
\begin{align*}
 q_{12}^{(\ell)}+ q_{22}^{(\ell)}=&\, \frac{1}{3}q_{21}^{(\ell)}q_{11}^{(\ell)}+\frac{1}{3}\left(q_{21}^{(\ell)}\right)^2,\\
 q_{14}^{(\ell)}+q_{24}^{(\ell)}=&\,(2q_{11}^{(\ell)}-q_{21}^{(\ell)})\left(-\frac{1}{9}q_{22}^{(\ell)}(q_{11}^{(\ell)}+q_{21}^{(\ell)})+\frac{1}{3}(q_{13}^{(\ell)}+q_{23}^{(\ell)})\right)\notag\\
 &+\frac{1}{3}q_{23}^{(\ell)}(q_{11}^{(\ell)}+q_{21}^{(\ell)}).
\end{align*}
Combining these with \eqref{eq:180117} we obtain the following characterisation: the monodromy about $w_{\ell}$ is trivial if and only if the
following system of 3 equations
\begin{align}
q_{12}^{(\ell)}=&\frac{1}{3}\left(\left(q_{11}^{(\ell)}\right)^2-q_{11}^{(\ell)}q_{21}^{(\ell)}+\left(q_{21}^{(\ell)}\right)^2\right)\label{finda21}\\
q_{22}^{(\ell)}=&\frac{1}{3}q_{11}^{(\ell)}\left(2q_{21}^{(\ell)}-q_{11}^{(\ell)}\right)\label{finda22}\\
q_{14}^{(\ell)}+q_{24}^{(\ell)}=&\frac{1}{3}q_{13}^{(\ell)}\left(2q_{11}^{(\ell)}-q_{21}^{(\ell)}\right)+q_{11}^{(\ell)}q_{23}^{(\ell)}\notag\\
&+\frac{1}{27} q_{11}^{(\ell)}\left(2q_{11}^{(\ell)}-q_{21}^{(\ell)}\right)\left(q_{11}^{(\ell)}-
2q_{21}^{(\ell)}\right)\left(q_{11}^{(\ell)}+q_{21}^{(\ell)}\right)\label{eq:180117-2}
\end{align}

In order to proceed further we write explicitly the coefficients $q$'s, which appear in the above equations,
in terms of the parameters of the opers \eqref{eq:3ass}
\begin{align*}
q_{10}^{\ell}&=3,\qquad q_{20}^{\ell}=3, \qquad
q_{11}^{(\ell)}=\frac{a_{11}^{(\ell)}}{w_\ell},\qquad q_{21}^{(\ell)}=\frac{a_{21}^{(\ell)}}{w_\ell},\\
q_{12}^{(\ell)}&=\frac{\bar{r}^1-a_{11}^{(\ell)}}{w_\ell^2}+\sum_{\substack{j=1\\j\neq\ell}}^N\left(\frac{3}{(w_\ell-w_j)^2}+\frac{a_{11}^{(j)}}{w_\ell(w_\ell-w_j)}\right),\qquad q_{22}^{(\ell)}=\frac{a_{22}^{(\ell)}-a_{21}^{(\ell)}}{w_\ell^2}\\
q_{13}^{(\ell)}&=\frac{a_{11}^{(\ell)}-2\bar{r}^1}{w_\ell^3}-\sum_{\substack{j=1\\j\neq \ell}}^N\left(\frac{6}{(w_\ell-w_j)^3}+\frac{a_{11}^{(j)}}{w_\ell(w_\ell-w_j)^2}+\frac{a_{11}^{(j)}}{w_\ell^2(w_\ell-w_j)}\right)\\
q_{23}^{(\ell)}&=\frac{\bar{r}^2+a_{21}^{(\ell)}-2a_{22}^{(\ell)}+w_\ell}{w_\ell^3}+\lambda w_\ell^k+\sum_{\substack{j=1\\j\neq \ell}}^N\left(\frac{3}{(w_\ell-w_j)^3}+\frac{a_{21}^{(j)}}{w_\ell(w_\ell-w_j)^2}+\frac{a_{22}^{(j)}}{w_\ell^2(w_\ell-w_j)}\right)\\
q_{14}^{(\ell)}&=\frac{3\bar{r}^1-a_{11}^{(\ell)}}{w_\ell^4}+\sum_{\substack{j=1\\j\neq \ell}}^N\left(\frac{9}{(w_\ell-w_j)^4}+\frac{a_{11}^{j}}{w_\ell(w_\ell-w_j)^3}+\frac{a_{11}^{j}}{w_\ell^2(w_\ell-w_j)^2}+\frac{a_{11}^{j}}{w_\ell^3(w_\ell-w_j)}\right)\\
q_{24}^{(\ell)}&=\frac{3a_{22}^{(\ell)}-a_{21}^{(\ell)}-3\bar{r}^2-2w_\ell}{w_\ell^4}+\lambda kw_\ell^{k-1}\\
&-\sum_{\substack{j=1\\j\neq \ell}}^N\left(\frac{9}{(w_\ell-w_j)^4}+\frac{2a_{21}^{(\ell)}}{w_\ell(w_\ell-w_j)^3}+\frac{a_{21}^{(\ell)}+a_{22}^{(\ell)}}{w_\ell^2(w_\ell-w_j)^2}+\frac{2a_{22}^{(\ell)}}{w_\ell^3(w_\ell-w_j)}\right).
\end{align*}

We notice that while equations \eqref{finda21} and \eqref{finda22} do not depend on $\la$,  equation \eqref{eq:180117-2} is a linear polynomial in $\lambda$. Since the trivial monodromy conditions must hold for any $\la$,
equation \eqref{eq:180117-2} consists of a pair of independent constraints:
both the constant part in $\la$ and the linear part in $\la$ are required to vanish independently.
The vanishing of the part of \eqref{eq:180117-2} which is linear in $\lambda$ reads:
$$q_{11}^{(\ell)}w_\ell^k-kw_\ell^{k-1}=0, \qquad \text{or} \qquad q_{11}^{(\ell)}=\frac{k}{w_\ell},$$
from which we obtain
\beq\label{a11k}
a_{11}^{(\ell)}=k,\qquad \ell=1,\dots,N.
\eeq
Making use of the explicit expression of the $q's$ in terms of the $a's$, as given above,
and denoting 
\beq
a_\ell=a_{21}^{(\ell)},\qquad \ell=1,\dots,N,
\eeq
from \eqref{finda22} we obtain
\beq\label{a22expr}
a_{22}^{(\ell)}=\frac{2}{3}(k+3)a_\ell-\frac{k^2}{3},\qquad \ell=1,\dots,N.
\eeq
Substituting \eqref{a11k} and \eqref{a22expr} into the expression for the $q$'s found above,
then from \eqref{finda21} we obtain \eqref{monodromy1}, while the vanishing of the constant
(in $\lambda$) coefficient of \eqref{eq:180117-2} is equivalent to \eqref{monodromy2}

We have thus arrived to the following result: an $\mf{sl}_3$ Quantum KdV oper is equivalent to a scalar third order differential operator
of the form (\ref{eq:a2scalaroper}) such that its coefficients satisfy the system of algebraic equations
\eqref{monodromy}.

\subsection{The dual representation. Formal adjoint operator}\label{sub:adjoint}
Before we proceed further with our analysis, and we construct solutions to the Bethe Ansatz equations,
we introduce a second representation of the algebra $\mf{sl}_3$. This is called
the second fundamental representation or dual representation, and we denote it by ${\bb{C}^3}^*$.
If $\lbrace\epsilon_1,\epsilon_2,\epsilon_3\rbrace$ is the standard basis of $\bb{C}^3$ as above, we denote by
 $\lbrace \epsilon^*_1,\epsilon^*_2,\epsilon^*_3 \rbrace$, the corresponding dual basis in $ {\bb{C}^3}^* $  such that
$\langle\epsilon^*_i,\epsilon_j\rangle=\delta_{ij}$. In these basis, the matrices representing the elements
$h_1,h_2,\theta^\vee,\rho^\vee,e_1,e_2,e_{\theta},f$ read
\begin{align*}& h_1=
\begin{pmatrix}
0 & 0 & 0\\
0 & 1 & 0\\
0 & 0 & -1
\end{pmatrix} \quad h_2=\begin{pmatrix}
1 & 0 & 0\\
0 & -1 & 0\\
0 & 0 & 0
\end{pmatrix}\quad 
\rho^\vee=\theta^\vee=
\begin{pmatrix}
1 & 0 & 0\\
0 & 0 & 0\\
0 & 0 & -1
\end{pmatrix} \\
& e_1=
\begin{pmatrix}
0 & 0 & 0\\
0 & 0 & 1\\
0 & 0 & 0
\end{pmatrix} \quad 
e_2=
\begin{pmatrix}
0 & 1 & 0\\
0 & 0 & 0\\
0 & 0 & 0
\end{pmatrix} \quad
e_{\theta}=
\begin{pmatrix}
0 & 0 & -1\\
0 & 0 & 0\\
0 & 0 & 0
\end{pmatrix} \quad
f=
\begin{pmatrix}
0 & 0 & 0\\
1 & 0 & 0\\
0 & 1 & 0
\end{pmatrix} 
\end{align*}
As we have discussed before, the general $\mf{sl}_3$ oper can be written the canonical form as
the connection $\mc{L}=\partial_z+f+v_1e_1+v_2e_{\theta}$, for an arbitrary pair of rational functions $v_1,v_2 \in K$.
In the dual representation, we thus have
$$\mc{L}=\partial_z+\begin{pmatrix}
0 & 0 & -v_2\\
1 & 0 & v_1\\
0 & 1 & 0
\end{pmatrix}. $$
We showed that in the standard representation $\bb{C}^3$ the  connection $\mc{L}$ is equivalent to the scalar third order operator \eqref{eq:scalarform}.  In the dual representation the same oper is equivalent to a different scalar operator, namely to its formal adjoint. Let $\psi^\ast=:\bb{C}\to {\bb{C}^3}^\ast$, with $\psi^\ast(z)=\psi^\ast_1(z)\epsilon^\ast_1+\psi^\ast_2(z)\epsilon^\ast_2+\psi^\ast_3(z)\epsilon^\ast_3$, satisfy $\mc{L}\psi^\ast=0$ in the dual representation. Then $\Psi^*(z):=\psi^*_3(z)$  satisfies the scalar ODE 
\begin{equation}\label{eq:scalardual}
 (-\partial_z^3+v_1 \partial_z + (v_2+v_1'))\Psi^*(z)=0 \; ,
\end{equation}
which is the formal adjoint of the equation \eqref{eq:scalarform}.
\\

The following standard isomorphisms (of $\mf{sl_3}-$modules) will be needed later to derive the Bethe Ansatz equations: $\bigwedge^2 \bb{C}^3 \cong {\bb{C}^3}^* $ and
$\bigwedge^2 {\bb{C}^3}^* \cong {\bb{C}^3}$. Explicitly,
\begin{gather}
 \imath(\epsilon_1\wedge \epsilon_2)=\epsilon_1^*,\quad \imath(\epsilon_1\wedge \epsilon_3)=\epsilon_2^*, \quad \imath(\epsilon_2 \wedge \epsilon_3)=\epsilon_3^* \\ \label{eq:isomorphisms}
 \imath(\epsilon_1^*\wedge \epsilon_2^*)=\epsilon_1,\quad  \imath(\epsilon_1^*\wedge \epsilon_3^*)=\epsilon_2, \quad\imath(\epsilon_2^* \wedge \epsilon_3^*)=\epsilon_3 
\end{gather}
The above isomorphisms imply that
if $\psi(z),\phi(z)$ are solutions of $\mc{L}\psi(z)=0$, for  $\psi:\bb{C} \to \bb{C}^3$ in the standard representation then
$\imath\big(\psi(z) \wedge \phi(z)\big)$ is a solution of the dual equation $\mc{L}\psi^*(z)=0$, with $\psi^*:\bb{C} \to {\bb{C}^3}^*$;
and conversely.

In the present paper we prefer to work with solutions of the equations in the scalar form \eqref{eq:scalarform} and \eqref{eq:scalardual}.
Recall that the solution of the equations in the scalar form is just the third component of the solution of the vector equation. If $\psi(z)=\psi_1(z)\epsilon_1+\psi_2(z)\epsilon_2+\psi_3(z)\epsilon_3$ and $\phi(z)=\phi_1(z)\epsilon_1+\phi_2(z)\epsilon_2+\phi_3(z)\epsilon_3$, then a simple
calculation shows that
$$\langle \imath (\psi\wedge\phi),\epsilon_3\rangle=Wr[\psi_3,\phi_3]$$
where $Wr[\cdot,\cdot] $ denotes the usual Wronskian $Wr[f(z),g(z)]=f(z)g'(z)-f'(z)g(z).$ Similarly, for   $\psi^\ast(z)=\psi^\ast_1(z)\epsilon^\ast_1+\psi^\ast_2(z)\epsilon^\ast_2+\psi^\ast_3(z)\epsilon^\ast_3$ and $\phi^\ast(z)=\phi^\ast_1(z)\epsilon^\ast_1+\phi^\ast_2(z)\epsilon^\ast_2+\phi^\ast_3(z)\epsilon^\ast_3$ we have
$$\langle \epsilon^\ast_3,\imath (\psi^*\wedge \phi^*)\rangle= Wr[\psi_3^*,\phi_3^*].$$
To prove the above relations, it is sufficient to note that from the matrix first order equations $\mc{L}\psi(z)=0,\mc{L}\psi^*(z)=0$ we obtain the identities $\psi_2(z)=-\psi_3'(z)$ and $\psi^*_2=-{\psi_3^*}'(z)$. We have thus shown that the Wronskian of two solutions of \eqref{eq:scalarform} satisfies \eqref{eq:scalardual}, and conversely the Wronskian of two solutions
of \eqref{eq:scalardual} satisfies \eqref{eq:scalarform}.

\subsection{Relation with previous works}
The ground state $\mf{sl}_3-$quantum KdV oper,  given by equation (\ref{eq:LGs}),  was also considered -- in the scalar form -- by Dorey and Tateo \cite{dota00}, and by Bazhanov, Hibberd and Khoroshkin \cite{bazhanov02integrable}, who wrote  the following third order scalar operator 
\beq\label{Lbhk}
\widetilde{L}(x,E)=\partial_x^3+\frac{\widetilde{w}_1}{x^2}\partial_x+\frac{\widetilde{w}_2}{x^3}+x^{3M}-E,
\eeq 
with $\widetilde{w}_1=\tilde{\ell}_1\tilde{\ell}_2+\tilde{\ell}_1\tilde{\ell}_3+\tilde{\ell}_2\tilde{\ell}_3-2$,
$\widetilde{w}_2=-\tilde{\ell}_1\tilde{\ell}_2\tilde{\ell}_3$ and where the $\tilde{\ell}_i$'s  are constrained by the equation
$\tilde{\ell}_1+\tilde{\ell}_2+\tilde{\ell}_3=3$. In addition, in our previous paper \cite{marava15} we considered the ground state oper in the following form
\beq\label{eq:L0x}
\mc{L}(x,E)=\partial_x+
\begin{pmatrix}
\ell_1/x & 0 & x^{3M}-E\\
1 & (\ell_2-\ell_1)/x & 0\\
0 & 1 & -\ell_2/x
\end{pmatrix}
\eeq
for arbitrary $\ell_1,\ell_2\in\bb{C}$ and $M>0$. We now show that the differential operators \eqref{eq:LGs}, \eqref{Lbhk}, and \eqref{eq:L0x} are equivalent under appropriate change of coordinates and Gauge transformations., once the parameters are correctly identified. To show that the differential operators \eqref{Lbhk} and \eqref{eq:L0x} are equivalent, we write
the operator \eqref{Lbhk} in the oper form
\beq\label{Lbhkmatrix}
\partial_x+
\begin{pmatrix}
0 & -\widetilde{w}_1/x^2 & \widetilde{w}_2/x^3+ x^{3M}-E\\
1 & 0 & 0\\
0 & 1 & 0
\end{pmatrix}.
\eeq
It is then a simple computation to show that \eqref{Lbhkmatrix} and \eqref{eq:L0x}
are Gauge equivalent if we set $\tilde{\ell}_1=-\ell_1+2$, $\tilde{\ell}_2=\ell_1-\ell_2+1$ and $\tilde{\ell}_3=\ell_2$.

Next we show the equivalence between \eqref{eq:L0x} and \eqref{eq:LGs}. 
As observed in \cite{FF11}, after the change of variable
\begin{align}\label{eq:xzchange}
z=\varphi(x)= \left(\frac{k+3}{3}\right)^{3} x^{\frac{3}{k+3}}, \quad
k=-\frac{3M+2}{1+M}
\end{align}
the operator \eqref{eq:L0x} reads 
\begin{align}\label{eq:L0}
\mc{L}_G(z,\la)&=\de_z+\begin{pmatrix}
r_1/z & 0 & z^{-2}+\lambda z^{k}\\
1 & (r_2-r_1)/z & 0\\
0 & 1 & -r_2/z
\end{pmatrix}
\end{align}
where $\la \in \bb{C}$ and $r_1,r_2 \in \bb{C}$ are defined by the relations
\begin{align}\label{eq:birel}
E=-\left(\frac{k+3}{3}\right)^{3(k+2)}\lambda, \quad \ell_i=\frac{3}{k+3} (r_i-1)+1,\, i=1,2 .
\end{align}
It is again a simple computation to show that the opers \eqref{eq:L0} and \eqref{eq:LGs} are Gauge equivalent
provided the coefficients $r^1,r^2,\bar{r}^1,\bar{r}^2$
satisfy the following relations
\beq\label{barr}
\begin{cases*}
\bar{r}^1=(r^1)^2-r^1r^2+(r^2)^2-r^1-r^2,\\
\bar{r}^2=r^1r^2(r^1-r^2)+r^2(2r^2-r^1-2).
\end{cases*}
\eeq
\subsection{Weyl group symmetry}
The parametrisation \eqref{barr} of $\bar{r}^1, \bar{r}^2$ in terms of $r^1$ and $r^2$
will be very convenient when discussing the behaviour of solutions of $\mc{L}(z,\la) \psi=0$
in a neighbourhood of $z=0$.
The Weyl group of $\mf{sl}_3$ -- which is isomorphic to the group of permutations of three elements, $S_3$ -- is a symmetry of the map
\eqref{barr}, once its action on the parameters $r^1,r^2$, which is called the \textit{dot action}, is properly defined:
\begin{equation}\label{eq:dotaction}
 \sigma\cdot  \begin{pmatrix} r^1\\r^2 \end{pmatrix} =  \begin{pmatrix} -r^2+2\\-r^1+2 \end{pmatrix} \, , \quad
 \tau\cdot \begin{pmatrix} r^1\\r^2 \end{pmatrix} =  \begin{pmatrix} -r^2+2\\r^1-r^2+1 \end{pmatrix} \; .
\end{equation}
We let the reader verify that $\sigma,\tau$ generate the group $S_3$ (in particular $\sigma^2=1,\tau^3=1$) and that
the above action is a symmetry of \eqref{barr}.
This phenomenon is studied in great detail and generality in \cite[Section 5]{mara18}.

\section{The Bethe Ansatz equations}\label{sec:BA}
In this Section we construct solutions of the Bethe Ansatz equations as generalised monodromy data of Quantum KdV opers,  $\mc{L}(z,\la)$. As proved in Section \ref{sec:Qopers}, these are opers of the form
\beq \label{eq:3ass4}
 \mc{L}(z,\la)=  \partial_z+
 \begin{pmatrix}
 0 & W_1 & W_2\\
 1 & 0 & 0\\
 0 & 1 & 0
 \end{pmatrix},
\eeq
where
\begin{subequations}\label{w1w24}
\begin{align}
W_1(z)&=\frac{\bar{r}^{1}}{z^{2}} +\sum_{j \in J}\left(\frac{3}{(z-w_j)^{2}}+\frac{k}{z(z-w_j)}\right) ,\\
W_2(z,\la)&=\frac{\bar{r}^{2}}{z^{3}} +\frac{1}{z^2}+\lambda z^k +\sum_{j \in J}\left(\frac{3}{(z-w_j)^{3}}+
\frac{a_j}{z(z-w_j)^{2}}+\frac{2(k+3)a_j-k^2}{3z^2(z-w_j)}\right),
\end{align}
\end{subequations}
and where $\{a_j,w_j\}_{j=1,\dots,N}$ satisfy the system of equations \eqref{monodromy}.
We follow
\cite{marava15,mara18} closely and the reader should refer to these papers for all missing proofs.
Any finite dimensional representation $V$ of $\mf{sl}_3$ defines the ODE 
$$\mc{L}(z,\la) \psi=0,\qquad \Psi : \bb{C} \to V.$$
Since the monodromy of $\mc{L}(z,\la)$ about $w_j$ is trivial for any $j$, then
the solutions of the above equation are, for fixed $\la$, analytic functions on the
universal cover of $\bb{C}^*$, minus the lift of the points $w_j, j \in 1\dots N$. We denote such a domain
by $\widehat{\bb{C}}$.
As it was originally observed by Dorey and Tateo, the appearance of the Bethe Ansatz
equations is due to a discrete symmetry which acts on both the variable $z$ and the parameter $\la$.
It is therefore necessary to consider solutions
$\psi(z,\la)$ as analytic functions of both variables $z$ and $\la$. More precisely for our purpose
$\psi(z,\cdot )$ is assumed to be an entire function of $\la$. We thus define a solution to be an analytic map
$\psi:\widehat{\bb{C}}\times \bb{C} \to V$ which satisfies
the equation $\mc{L}(z,\la) \psi(z,\la)=0$ for every $(z,\la)$.

The space of solutions, which we denote by $V(\la)$, is an infinite dimensional vector space which, as we showed in \cite{mara18}, is simply isomorphic to $V \otimes O_{\la}$, where $O_{\la}$ is the ring of entire functions of the variable $\la$.
This means that an $O_{\la}-$basis of the space of solutions has cardinality $\dim V$.
\subsection{Twisted opers} Let $\hat{k}=-k-2$, so that $0<\hat{k}<1$. For any $t \in \bb{R}$ we define the twisted operator and twisted solution:
\begin{align}\label{eq:twistedoper}
 & \mc{L}^t(z,\la):=\mc{L}(e^{2i \pi t}z,e^{2i \pi t \hat{k}}\la) \\ \label{eq:twistedsolution}
 & \psi_{t}(z,\la)=e^{2 i \pi t \rho^\vee}\psi(e^{2\pi i t}z, e^{2 \pi i t \hat{k}}\la)
\end{align}
Taking into account the oper change of variables \cite{mara18}, then from \eqref{eq:3ass4} we explicitly have
\begin{equation*}
 \mc{L}^t(z,\la)=\partial_z+f+ e^{4 \pi i t}W_1(e^{2 \pi i t}z) e_1 + e^{6 \pi i t} W_2(e^{2 \pi it} z, e^{2 \pi i t \hat{k}} \la)e_\theta,
\end{equation*}
and one easily see that the function $\psi_t(z,\la)$ satisfies $\mc{L}^t(z,\la)\psi_t(z,\la)=0$. A crucial property of the oper \eqref{eq:3ass4} is the following Dorey-Tateo discrete symmetry:
\begin{equation}\label{eq:doreytateosymmetry}
 \mc{L}^{t=1}(z,\la)=\mc{L}(z,\la) \; ,
\end{equation}
which leads us to consider the following ($O_{\la}-$linear) monodromy operator
\begin{equation}\label{eq:Moperator}
 M: V(\la) \to V(\la), \qquad M(\psi(z,\la))=e^{2i \pi \rho^\vee}\psi(e^{2\pi i} z,e^{2 \pi i \hat{k}} \la) \; .
\end{equation}

In the case $\mf{sl}_3$, we just need to consider the equations $\mc{L}^t(z,\la)\psi=0$ for the standard representation and its dual.
More precisely, the standard representation at $0$ twist, and the dual representation at twist $t=\frac12$
\begin{align}\label{eq:matbethe1}
& \mc{L}(z,\la) \psi(z,\la)=0, \qquad \psi: \widehat{\bb{C}}\times \bb{C} \to \bb{C}^3\\ \label{eq:matbethe2}
& \mc{L}^{\frac12}(z,\la) \psi^*(z,\la)=0, \qquad \psi^*: \widehat{\bb{C}}\times \bb{C} \to {\bb{C}^3}^*
\end{align}
By a (slight abuse of notation) we denote $\bb{C}^3(\la)$ the space of solutions of the first equation, and by ${\bb{C}^3}^*(\la)$ the space
of solutions of the latter equations, as well as the solutions of the same equations in the equivalent scalar form. 
\begin{align}\label{eq:scalarbethe1}
& \big(\partial_z^3 - W_1(z) \partial_z+ W_2(z,\la) \big) \Psi(z,\la)=0 \\ \label{eq:scalarbethe2}
&  \big(\partial_z^3 - W_1(-z) \partial_z+ W_2(-z,e^{\pi i \hat{k}}\la)-W_1'(-z) \big) \Psi^*(z,\la)=0
\end{align}
Since the solution of the equations in the scalar form is the third component of a solution of the equation in the matrix form, and since $\rho^\vee \epsilon_3=-\epsilon_3,\rho^\vee \epsilon^*_3=-\epsilon^*_3$, the twist for solutions of the above scalar ODEs is defined as follows
\begin{equation*}
 \Psi_t(z,\la)=e^{-2 i \pi t}\Psi(e^{2\pi i t}z, e^{2 \pi i t \hat{k}}\la) \, , \quad
 \Psi^*_t(z,\la)=e^{-2 i \pi t}\Psi^*(e^{2\pi i t}z, e^{2 \pi i t \hat{k}}\la)
\end{equation*}
Equation (\ref{eq:scalarbethe2}) is the adjoint equation to \eqref{eq:scalarbethe1} twisted by $t=\frac12$; and conversely, equation (\ref{eq:scalarbethe1}) is the adjoint equation to \eqref{eq:scalarbethe2} twisted by $t=\frac12$. As we recalled
in Subsection \ref{sub:adjoint}, the Wronskian of two solutions of a scalar ODE solves the adjoint equation. 
It follows that
\begin{enumerate}
 \item If $\Psi(z,\la),\Phi(z,\la) \in \bb{C}^3(\la)$
 then 
 $$Wr[\Psi_{-\frac12}(z,\la),\Phi_{\frac12}(z,\la)] \in {\bb{C}^3}^*(\la),$$
 \item If  $\Psi^*(z,\la),\Phi^*(z,\la) \in {\bb{C}^3}^*(\la)$, then 
$$Wr[\Psi^*_{-\frac12}(z,\la),\Phi^*_{\frac12}(z,\la)] \in \bb{C}^3(\la).$$
\end{enumerate}

\subsection{The eigenbasis of the monodromy operator. Expansion at $z=0$}
The point $z=0$ is a regular singularity for the equations \eqref{eq:scalarbethe1}, \eqref{eq:scalarbethe2}, but it is also
a branch point of the potential $W_2$, because of the term $\la z^k$. It follows that the standard Frobenius series
cannot provide solution of the above equations at $z=0$. A generalised Frobenius series,
introduced in \cite{mara18}, does however the job. The latter is defined as
\begin{equation}\label{eq:genfrobenius}
 \Phi^{(\beta)}(z,\la)=z^{\beta} \sum_{m\geq n \geq 0} c_{m,n} z^m \zeta^n \, , \;  c_{0,0}=1 \, , 
 \quad \zeta=\la z^{-\hat{k}} \; .
\end{equation}
where the indices $\beta$ are computed as in the standard Frobenius method:
if the equation reads 
$$\left(\partial_z^3+\frac{a+o(1)}{z^2}\partial_z+\frac{b+o(1)}{z^3}\right)\Psi(z)=0,$$ 
the indices are the roots of the indicial polynomial $P(\beta)=\beta^3-3 \beta^2+(2+a)\beta+b$. The following facts are proved in \cite[Proposition 5.1]{mara18}. For every finite dimensional representation $V$ of $\mf{sl}_3$, and
under some genericity assumptions \footnote{The genericity assumptions imply that the monodromy operator  $M$ is diagonal
and no logarithmic terms appear in the generalised Frobenius series.} on the triple $(\hat{k},\bar{r}^1,\bar{r}^2)$, we have:
\begin{enumerate}
 \item The series (\ref{eq:genfrobenius}) converges to a solution $\Phi^{(\beta)}(z,\la) \in V(\la)$.
 \item $M\Phi^{(\beta)}(z,\la)=e^{2 \pi i \beta} \Phi^{(\beta)}(z,\la)$, where $M$ is the monodromy operator defined in \eqref{eq:Moperator}.
 \item The collection of the solutions $\Phi^{(\beta)}(z,\la)$ for all indices $\beta$ forms an $O_{\la}-$basis of $V(\la)$.
\end{enumerate}

In the cases under our study,
namely equations \eqref{eq:scalarbethe1} and \eqref{eq:scalarbethe2},
the indicial polynomials are, respectively, given by
\begin{align*}
P(\beta)&=\beta^3-3 \beta^2+(2-\bar{r}^1)\beta+\bar{r}^2,\\
P^*(\beta)&=\beta^3-3\beta^2+(2-\bar{r}^1)+2\bar{r}^1-\bar{r}^2. 
\end{align*}
Using \eqref{barr}, then we obtain the factorizations
\begin{align*}
P(\beta)&=(\beta-r^2)(\beta-1+r^2-r^1)(\beta-2+r^1),\\ 
P^*(\beta)&=(\beta-r^1)(\beta-1+r^1-r^2)(\beta-2+r^2), 
\end{align*}
so that the indices are given by
\begin{subequations}\label{eq:indices}
\begin{gather}
\beta_1=r^2,\quad \beta_2=r^1-r^2+1,\quad \beta_3=-r^1+2 \\ 
\beta_1^*=-r^2+2,\quad \beta_2^*=r^2-r^1+1,\quad \beta_3^*=r^1.
\end{gather}
\end{subequations}
We denote by 
\begin{subequations}\label{phis}
\begin{gather}
\lbrace \Phi^{(\beta_1)}(z,\la),\Phi^{(\beta_2)}(z,\la),\Phi^{(\beta_3)}(z,\la),\rbrace\label{phis1}\\
\lbrace \Phi^{(\beta^\ast_1)}(z,\la),\Phi^{(\beta^\ast_2)}(z,\la),\Phi^{(\beta^\ast_3)}(z,\la)\rbrace\label{phis2},
\end{gather}
\end{subequations}
the corresponding solutions of \eqref{eq:scalarbethe1} and \eqref{eq:scalarbethe2} respectively.
Recall that the Weyl group acting by the dot action \eqref{eq:dotaction} on $r^1,r^2$, provides a group of symmetries of $\bar{r}^1,\bar{r}^2$, hence it leaves the indicial polynomial invariant, permuting its roots\footnote{Many authors fix $r^1,r^2$ by imposing the conditions $\Re \beta_1 >\Re \beta_2> \Re \beta_3$ , or equivalently
$\Re \beta^*_3 >\Re \beta^*_2> \Re \beta^*_1$ \cite{bazhanov02integrable,dorey07,marava15}}.
The (induced) action of the generators $\sigma,\tau$ of the Weyl group, see \eqref{eq:dotaction}, on the indices
\eqref{eq:indices} is provided by the following permutations:
\begin{subequations}\label{inducedaction}
\begin{align} 
&\,\sigma (\beta_i)=\beta_{\sigma(i)}, \quad\, \tau (\beta_i)=\beta_{\tau(i)}, \qquad i=1,2,3\\
&\sigma (\beta^*_i)=\beta^*_{\sigma(i)}, \quad \tau (\beta^*_i)=\beta_{\tau(i)}, \qquad i=1,2,3
\end{align}
\end{subequations}
where
\beq\label{eq:permutations}
\sigma(1,2,3):=(3,2,1),\qquad \tau(1,2,3):=(2,3,1) .
\eeq
Comparing the asymptotic behaviour at $z=0$, we deduce the following $6$ quadratic identities
among the (properly normalised) $\Phi^{(\beta)}$'s and $\Phi{\beta^*}$'s.
Let $s \in S_3$, then (we can find a normalisation of the solutions $\Phi^{\beta_i},\Phi^{\beta_i^*}$ such that):
\begin{subequations}\label{eq:PhiPhi}
\begin{align}
 & Wr[\Phi_{\mp\frac12}^{(\beta_{s(1)})},\Phi_{\pm \frac12}^{(\beta_{s(2)})}]= (-1)^{p(s)} e^{ \pm i \pi  s(\gamma) }
 {\Phi}^{(\beta_{s(3)}^*)},\\ 
 & Wr[ {\Phi_{\mp\frac12}^{(\beta^*_{s(3)})}},{\Phi_{\pm\frac12}^{(\beta^*_{s(2)})}}]= (-1)^{p(s)}  e^{\pm i \pi s(\gamma^*)}  \Phi^{(\beta_{s(1)})}, 
\end{align}
\end{subequations}
where $p(s)$ is the parity of $s\in S_3$, and  where 
\beq\label{sgamma} 
s(\gamma)=\beta_{s(2)}-\beta_{s(1)},\qquad s(\gamma^*) = \beta^*_{s(2)}-\beta^*_{s(3)},
\eeq
with the $\beta_{s(i)}$ and $\beta^\ast_{s(i)}$ defined by the relations (\ref{eq:indices},\ref{inducedaction}).

\subsection{Sibuya solutions. Expansion at $z=\infty$}
We let $q(z,\la)$ be the Puiseaux series of $\big(z^{-2}(1+\la z^{-\hat{k}})\big)^{\frac13}$ truncated after terms of $z^{-1}$, 
and $S(z,\la)$ be its primitive
\begin{equation}\label{eq:qzl}
q(z,\la)= z^\frac{-2}{3}\big(1+ \sum_{l=0}^{\lfloor \frac{1}{3 \hat{k}} \rfloor} c_l \la^l z^{-l \hat{k}} \big),\qquad
S(z,\la)= \int^z q(y,\la) dy,
\end{equation}
where $c_l$ are the coefficients of Taylor series expansion at $y=0$ of $(1-y)^{\frac13}$, and
$\int^z y^l dl=\frac{z^{l+1}}{l+1}, l \neq -1 $, $\int^z \frac{1}{y}=\log z$.

The Sibuya, or subdominant, solution of the equations \eqref{eq:scalarbethe1}, \eqref{eq:scalarbethe2} is uniquely
defined by the following asymptotics
\begin{subequations}\label{eq:sibuya}
\begin{align}
\Psi(z,\la)&=z^{\frac23}e^{-S(z,\la)} \big( 1+ o(1) \big)  , \quad \mbox{ as } z \to +\infty,  \\ 
\Psi^*(z,\la)&=z^{\frac23}e^{-S(z,\la)}  \big( 1 + o(1) \big), \quad \mbox{ as } z \to +\infty. 
\end{align}
\end{subequations}
Moreover we have that
\begin{subequations}\label{eq:sibuyader}
\begin{align}
\Psi'(z,\la)&=-e^{-S(z,\la)} \big( 1+ o(1) \big), \quad \mbox{ as } z \to +\infty, \\
 {\Psi^*}'(z,\la)&=-e^{-S(z,\la)}  \big(1+ o(1) \big) , \quad \mbox{ as } z \to +\infty.
\end{align}
\end{subequations}
The Sibuya solutions $\Psi,\Psi^*$ satisfy the following properties
\begin{itemize}
\item It is the solution (unique up to a multiplicative constant) with the fastest decrease as
$z\to + \infty$. 
 \item The asymptotic formulas \eqref{eq:sibuya} hold true on the sector
 $|\arg{z}|\leq \pi +\e$, for some $\e>0$ \cite{marava15}. In other words, if we continue analytically $\Psi(z,\la),\Psi^*(z,\la)$ as well as the functions $q(z,\la)$ and $S(z,\la)$ past the negative real semi-axis, the asymptotic formulas still hold.
 \item The solutions $\Psi(z,\la),\Psi^*(z,\la)$ are entire functions of $\la$, i.e. $\Psi(z,\la) \in \bb{C}^3(\la)$ and
$\Psi^*(z,\la) \in {\bb{C}^3}^*(\la)$.
\item Finally, and most importantly, the solutions $\Psi(z,\la),\Psi^*(z,\la)$ satisfy the so-called $\Psi$-system
\begin{subequations}\label{eq:psisystem}
\begin{align}\label{eq:psisystem1}
& Wr[\Psi_{-\frac12}(z,\la),\Psi_{\frac12}(z,\la)])=\Psi^*(z,\la)  \\ \label{eq:psisystem2}
& Wr[\Psi^*_{-\frac12}(z,\la),\Psi^*_{\frac12}(z,\la)])=\Psi(z,\la)
\end{align}
\end{subequations}
The latter identities can be checked by comparing the asymptotic expansion of the left and right
hand side as $z\to +\infty$.
\end{itemize}
The $\Psi$-system is the last necessary ingredient to construct solutions of the Bethe Ansatz equations.

\subsection{$Q\widetilde{Q}$ system and the Bethe Ansatz}
As we have shown, the solutions $\lbrace \Phi^{(\beta_1)}(z,\la)$,$\Phi^{(\beta_2)}(z,\la)$,
$\Phi^{(\beta_3)}(z,\la)\rbrace$ obtained in \eqref{phis1} provide an $O_{\la}$ basis of $\bb{C}^3(\la)$, and
the Sibuya solution $\Psi(z,\la)$ belongs to the same space.
It follows that there exists a unique triplet of entire functions $Q_{i}(\la) \in O_{\la}$, for 
$i=1,2,3$, such that
\begin{subequations}\label{eq:QQdualexp}
\begin{equation}\label{eq:Qexp}
\Psi(z,\la)=Q_{1}(\la)\Phi^{(\beta_1)}(z,\la)+Q_{2}(\la)\Phi^{(\beta_2)}(z,\la)+
Q_{3}(\la)\Phi^{(\beta_3)}(z,\la) \; .
\end{equation}
Similarly, we have that
\begin{equation}\label{eq:Qdualexp}
\Psi^*(z,\la)=Q_{1}^*(\la)\Phi^{(\beta^*_1)}(z,\la)+{Q_{2}^*}(\la)\Phi^{(\beta^*_2)}(z,\la)+
{Q_{3}^*(\la)}\Phi^{(\beta^*_3)}(z,\la) \; ,
\end{equation}
\end{subequations}
for a unique triplet of entire functions $Q_{i}^*(\la) \in O_{\la}$, with  $i=1,2,3$. Substituting the expansions \eqref{eq:QQdualexp} in the $\Psi$-system \eqref{eq:psisystem}
and making use of the relations \eqref{eq:PhiPhi} we obtain the following quadratic relations among the coefficients
$Q$'s and $Q^*$'s, which is known as $Q\widetilde{Q}-$system. For each $s\in S_3$ we have
\begin{subequations}\label{eq:QQtilde}
\begin{align} 
(-1)^{p(s)}{Q_{s(3)}^*}(\la)=&\,\, e^{i \pi s(\gamma) } Q_{s(1)}(e^{-i \pi \hat{k}} \la)Q_{s(2)}(e^{i \pi \hat{k}} \la)\notag\\
& -e^{-i \pi s(\gamma)}  Q_{s(1)}(e^{i \pi \hat{k}} \la) Q_{s(2)}(e^{-i \pi \hat{k}} \la)\\
(-1)^{p(s)}{Q_{s(3)}}(\la)=& \,\, e^{i \pi s(\gamma^*) } {Q_{s(1)}^*}(e^{-i \pi \hat{k}} \la){Q_{s(2)}}(e^{i \pi \hat{k}} \la)\notag\\
 &-e^{-i \pi s(\gamma^*)}  {Q_{s(1)}^*}(e^{i \pi \hat{k}} \la) {Q_{s(2)}^*}(e^{-i \pi \hat{k}} \la).
\end{align}
\end{subequations}
where $p(s)$ is the parity of $s$, and the phases $s(\gamma),s(\gamma^*)$ are defined in \eqref{sgamma}. 
\begin{remark}
System \eqref{eq:QQtilde} was shown by Frenkel and Hernandez \cite{fh16} to be a universal system of relations in the commutative Grothendieck ring $K_0(\mc{O})$ of the category $\mc{O}$ of representations of the Borel subalgebra of the quantum affine algebra $U_q(\widehat{\mf{sl}_3})$.
\end{remark}
Finally, the Bethe Ansatz equations is a pair of functional relations
for each one of the six pairs of $Q$ functions of the form $\lbrace Q_{s(1)} Q^*_{s(3)} \rbrace $, $s \in S_3$.
Let $\la_{s}$ denote an arbitrary zero of the function $Q_{s(1)}$,and ${\lambda^*_s}$ an arbitrary
zero of $Q_{s(3)}^*$. Evaluating the above relations at $e^{\pm i \pi \hat{k}}\la_s$ we obtain the Bethe Ansatz equations
\begin{align*}
& -e^{2 i \pi s(\gamma)}\frac{ Q_{s(1)}(e^{2 i \pi \hat{k}} \la_{s})}{Q_{s(1)}(e^{-2 i \pi \hat{k}} \la_{s})} =
\frac{ Q_{s(3)}^* (e^{ i \pi \hat{k}}\la_{s})}{ Q_{s(3)}^* (e^{ i \pi \hat{k}}\la_{s})} \\ \label{eq:bethe1}
&  -e^{2 i \pi s(\gamma^*)}\frac{ Q_{s(3)}^*(e^{2 i \pi \hat{k}} \la_{s}^*)} {Q_{s(3)}^*(e^{-2 i \pi \hat{k}} \la_{s}^*)} =
\frac{ {Q_{s(1)}} (e^{ i \pi \hat{k}}\la_{s}^*)}{ {Q_{s(1)}}(e^{ -i \pi \hat{k}}\la_{s}^*)}
\end{align*}
It is believed that each one of the $6$ Bethe Ansatz equations is strong enough to characterise all of the $Q$'s and $Q^*$'s,
by means of the so-called Destri-De Vega equations.

\section{Quantum Boussinesq Model}\label{sec:QBM}
The quantum Boussinesq model has been described in great detail by Bazhanov, Hibberd and Khoroshkin in
\cite{bazhanov02integrable}, from which the notation of the present section is taken and to which we refer for further details. The model is defined by considering a highest weight representation $\mc{V}_{\Delta_2,\Delta_3}$ of the 
Zamolodchikov's $\mc{W}_3$-algebra \cite{za85}, and it is characterized by $4$ parameters: the central charge $-\infty<c<2$, the highest weight $(\Delta_2,\Delta_3)\in\bb{C}^2$, and the
spectral parameter $\mu\in\bb{C}$. For generic values of $c,\Delta_2,\Delta_3$, the representation $\mc{V}_{\Delta_2,\Delta_3}$ is irreducible, a condition we assume from now on. Let ${\bf L}_n,{\bf W}_n$, $n\in\bb{Z}$, denote the generators of the $\mc{W}_3$ algebra as in \cite[Section 2]{bazhanov02integrable}. The highest weight fixes a ground state $|\Delta_2,\Delta_3\rangle\in \mc{V}_{\Delta_2,\Delta_3}$, satisfying ${\bf L}_n|\Delta_2,\Delta_3\rangle={\bf W}_n|\Delta_2,\Delta_3\rangle=0$ for $n>0$, and 
$${\bf L}_0|\Delta_2,\Delta_3\rangle=\Delta_2|\Delta_2,\Delta_3\rangle\qquad {\bf W}_0|\Delta_2,\Delta_3\rangle=\Delta_3|\Delta_2,\Delta_3\rangle.$$
The $\mc{W}_3$-module $\mc{V}_{\Delta_2,\Delta_3}$ admits the level decomposition
$$\mc{V}_{\Delta_2,\Delta_3}=\bigoplus_{N=0}^\infty \mc{V}_{\Delta_2,\Delta_3}^{(N)},\qquad {\bf L}_0\mc{V}_{\Delta_2,\Delta_3}^{(N)}=(\Delta_2+N)\mc{V}_{\Delta_2,\Delta_3}^{(N)}$$
The ground state $|\Delta_2,\Delta_3\rangle$ has level zero, the higher states are obtained by the action of products of the lowering operators ${\bf L}_n,{\bf W}_n$, $n<0$. More precisely, let $\lbrace \nu_1\dots,\nu_l,\bar{\nu}_1\dots,\bar{\nu}_k \rbrace $, with $\nu_j,\bar{\nu}_j \in \bb{N}$, be a bicoloured integer partition of the integer $N$, namely $\nu_j\leq \nu_{j+1}$, $\bar{\nu}_j\leq \bar{\nu}_{j+1}$ and $\sum_j \nu_j+ \sum_j \bar{\nu}_j =N$;
to any such a partition one associates a state of level $N$ by the formula
$\prod_{j} {\bf L}_{-\nu_j} \prod_{j} {\bf W}_{-\bar{\nu}_j} |0\rangle$.

The integrable structure of the quantum Boussinesq model can be conveniently encoded in the so-called ${\bf{Q}}-$operators \cite[Section 2]{bazhanov02integrable}, from which the quantum integrals of motion of the model can be obtained. The ${\bf{Q}}-$operators are more precisely operator-valued functions ${\bf Q}_i(t)$, $\overline{{\bf Q}}_i(t)$, $i=1,2,3$, depending on the parameter $t=\mu^3$, where $\mu$ is the spectral parameter of the quantum model\footnote{the spectral parameter $\mu$ is denoted $\lambda$ in \cite{bazhanov02integrable}}. The level subspaces $\mc{V}_{\Delta_2,\Delta_3}^{(N)}$ are invariant with respect to the action of the ${\bf Q}-$operators,
$${\bf Q}_i(t):\mc{V}_{\Delta_2,\Delta_3}^{(N)}\to \mc{V}_{\Delta_2,\Delta_3}^{(N)},\qquad \overline{{\bf Q}}_i(t):\mc{V}_{\Delta_2,\Delta_3}^{(N)}\to \mc{V}_{\Delta_2,\Delta_3}^{(N)},$$
and in particular (for $N=0$), the ground state $|\Delta_2,\Delta_3\rangle$ is an eigenvector for the ${\bf Q}-$operators:
\begin{align*}
{\bf Q}_i(t)|\Delta_2,\Delta_3\rangle&=P_i^{(vac)}(t)|\Delta_2,\Delta_3\rangle,\\
\overline{{\bf Q}}_i(t)|\Delta_2,\Delta_3\rangle&=\overline{P}_i^{(vac)}(t)|\Delta_2,\Delta_3\rangle.
\end{align*}
As proved in \cite[Section 5]{bazhanov02integrable}, the $\bf{Q}-$operators (and therefore their eigenvalues) satisfy the system of quadratic relations
\begin{subequations}\label{bhkQQsystem}
\begin{align}
c_1 \overline{{\bf Q}}_1(t)&={\bf Q}_2(qt){\bf Q}_3(q^{-1}t)-{\bf Q}_3(qt){\bf Q}_2(q^{-1}t)\\
c_1 {\bf Q}_1(t)&=\overline{{\bf Q}}_3(qt)\overline{{\bf Q}}_2(q^{-1}t)-\overline{{\bf Q}}_2(qt)\overline{{\bf Q}}_3(q^{-1}t)\\
c_2 \overline{{\bf Q}}_2(t)&={\bf Q}_3(qt){\bf Q}_1(q^{-1}t)-{\bf Q}_1(qt){\bf Q}_3(q^{-1}t)\\
c_2 {\bf Q}_2(t)&=\overline{{\bf Q}}_1(qt)\overline{{\bf Q}}_3(q^{-1}t)-\overline{{\bf Q}}_3(qt)\overline{{\bf Q}}_1(q^{-1}t)\\
c_3 \overline{{\bf Q}}_3(t)&={\bf Q}_1(qt){\bf Q}_2(q^{-1}t)-{\bf Q}_2(qt){\bf Q}_1(q^{-1}t)\\
c_3 {\bf Q}_3(t)&=\overline{{\bf Q}}_2(qt)\overline{{\bf Q}}_1(q^{-1}t)-\overline{{\bf Q}}_1(qt)\overline{{\bf Q}}_2(q^{-1}t),
\end{align}
\end{subequations}
where 
\begin{subequations}\label{parametersbhk}
\begin{align}
q&=e^{i\pi g},\\
c_1&=e^{i\pi (p_1-\sqrt{3}p_2)}-e^{-i\pi (p_1-\sqrt{3}p_2)},\\
c_2&=e^{-2i\pi p_1}-e^{2i\pi p_1},\\
c_3&=e^{i\pi (p_1+\sqrt{3}p_2)}-e^{-i\pi (p_1+\sqrt{3}p_2)},
\end{align}
\end{subequations}
and the parameter $g,p_1,p_2$ are related to $c,\Delta_2,\Delta_3$ by the identities \cite[Section 3]{bazhanov02integrable}
\beq\label{cdeltas}
c=50-24(g+g^{-1}),\quad \Delta_2=\frac{p_1^2+p_2^2}{g}+\frac{c-2}{24},\quad \Delta_3=\frac{2p_2(p_2^2-3p_1^2)}{(3g)^{3/2}}.
\eeq

\subsection{From \eqref{eq:QQtilde} to \eqref{bhkQQsystem}} We now prove that the $Q\widetilde{Q}$ system
\eqref{eq:QQtilde} and the system \eqref{bhkQQsystem} are equivalent. As a by-product we deduce the
explicit relations \eqref{parametersintro} among the parameters of the opers, $\bar{r}^1,\bar{r}^2,k,\la$, and
the parameters of the quantum theory,
$c,\Delta_2,\Delta_3,\mu$. More precisely, we derive (\ref{eq:ua},\ref{eq:ub},\ref{eq:uc}) while
\eqref{eq:corrparameters} can be found in \cite{bazhanov02integrable}.

Let $Q_i$, $Q^\ast_i$. $i=1,2,3$ be the functions defined by the expasions \eqref{eq:QQdualexp}, satisfying the $Q\widetilde{Q}-$system
\eqref{eq:QQtilde}. Assume that $Q_i(0)\neq 0$ and $Q^\ast_i(0)\neq 0$, $i=1,2,3$. Recall the definition of the indices $\beta_i$, $\beta^\ast_i$, $i=1,2,3$ as given in \eqref{eq:indices}. Then, a direct calculation shows that the functions
$$P_{i}(t)=t^{\beta_i}\frac{Q_i(t)}{Q_i(0)},\qquad P^\ast_{i}(t)=t^{\beta^\ast_i}\frac{Q^\ast_i( t)}{Q^\ast_i(0)},\qquad i=1,2,3.$$
satisfy \eqref{bhkQQsystem}, with the parameters $g,p_1,p_2$ appearing in \eqref{parametersbhk}
related to the parameters $\hat{k},r^1,r^2$ by the relations.
\beq\label{lastparameters}
g=1-\hat{k}=k+3,\quad p_1=\frac{r^1}{2}+\frac{r^2}{2}-1,\qquad p_2=\frac{\sqrt{3}}{2}(r^1-r_2).
\eeq
Substituting the above equation into \eqref{cdeltas} and using \eqref{barr} we obtain (\ref{eq:ua},\ref{eq:ub},\ref{eq:uc}).

\def\cprime{$'$} \def\cprime{$'$} \def\cprime{$'$} \def\cprime{$'$}
  \def\cprime{$'$} \def\cprime{$'$} \def\cprime{$'$} \def\cprime{$'$}
  \def\cprime{$'$} \def\cprime{$'$} \def\cydot{\leavevmode\raise.4ex\hbox{.}}
  \def\cprime{$'$} \def\cprime{$'$} \def\cprime{$'$}

\end{document}